# Sub-cycle doublon-holon dynamics in one-dimensional Mott insulators revealed by two-color high-harmonic spectroscopy

Lance Hatch[1], Aditya Verma[1], Eric Schultz[2], Hanjun Yang[1], Priscila Rosa[3,4], Genda Gu[5], Igor Zaliznyak[5], Giulio Vampa[6], Laimei Nie[2], and Hanzhe Liu[1*]

[1]Department of Chemistry, Purdue University, West Lafayette, Indiana 47907, USA
[2]Department of Physics and Astronomy, Purdue University, West Lafayette, Indiana 47907, USA
[3]Los Alamos National Laboratory, Los Alamos, New Mexico 87545, USA
[4]Department of Physics, Colorado State University, Fort Collins, CO 80523, USA
[5]Condensed Matter Physics and Materials Science Division, Brookhaven National Laboratory, Upton, New York 11973, USA
[6]Joint Attosecond Science Laboratory (JASLab), National Research Council of Canada and University of Ottawa, Ottawa, ON, Canada

**ABSTRACT**. Solid-state high-harmonic spectroscopy is becoming an emerging tool for probing nonequilibrium many-body dynamics. Yet, direct measurements of strongly driven, sub-optical-cycle dynamics in correlated materials during high-harmonic emission remain largely unexplored. Here, we measure high-harmonic emission chirp in a prototypical one-dimensional Mott insulator, which encodes strongly driven doublon-holon dynamics at sub-optical-cycle timescales. We observe a positive chirp for above band gap harmonics, indicating that high harmonics are dominated by doublon-holon recombinations. We further show a harmonic order-dependent dephasing, which can be understood through different doublon-holon excursion distance associated with each harmonic. These results reveal coherent doublon-holon dynamics and their ultrafast dephasing in Mott insulators, which is relevant to other nonequilibrium light-induced phenomena, such as Floquet engineering.

High-harmonic generation is a strong field nonlinear optical phenomenon and the key process for attosecond technologies [1,2]. It was first discovered in atoms [3–5], and later applied to molecules [6–12], liquids [13–16] and various solids [17–25]. In the gas phase, the emission of high harmonics is successfully described by the semiclassical three-step model [26–28]. In this description, high harmonics are initiated through tunneling ionization of a valence electron, followed by laser acceleration and recombination. A similar approach in the energy-momentum space has been developed to describe the electron-hole dynamics in wide band gap semiconductors [29–32].

One of the key results from these studies is that the high-harmonic emission is chirped, meaning different harmonic orders are emitted at different times within a single optical cycle of the driving field [33,34]. This chirp, often termed as attochirp in literature [35–38], naturally arises from the semiclassical model of high-harmonic generation. In the semiclassical description, each high harmonic is associated with a different and unique electron trajectory, each exhibiting different electron energies upon recombination [33,34]. In solids, these distinct electron energies result from the traversal of different distances within the crystal lattice, ultimately leading to varying recombination times within a single optical cycle [29,30,39]. Measuring attochirp is important for both understanding high-harmonic generation mechanisms and probing sub-cycle electronic dynamics. For example, quantifying attochirp in atoms and molecules has enabled the reconstruction of electron wavefunctions and real-space attosecond charge migration pathways [40–44]. In solids, attochirp measurements lead to the all-optical probing of band structures and topological properties [23–25,45–49].

Recently, high-harmonic spectroscopy has been extended to strongly-correlated materials [50–58]. It has been shown that high-harmonic generation is sensitive to various phase transitions and correlated many-body effects [50–58]. Importantly, theoretical studies have shown that sub-cycle electronic responses could reveal correlated many-body dynamics [59–69]. Yet, in most experimental studies, only cycle-averaged high-harmonic spectra were measured. Direct experimental measurements of the sub-cycle correlated electron dynamics during high-harmonic generation in strongly correlated materials remain largely unexplored.

Here, we report experimental measurements of high-harmonic chirp in a strongly-driven one-dimensional (1D) Mott insulator $Ca_{0.4}Sr_{0.6}CuO_2$, which reveals sub-cycle correlated doublon-holon motion. We observe chirp from above-gap high harmonics that is consistent with high-harmonic generation in Mott insulators being governed by doublon-holon recombination. In addition to observing sub-cycle doublon-holon dynamics, we measure high-harmonic temporal distribution within the multi-cycle driving field. We observe that higher harmonics are more suppressed than lower harmonics in the later part of the driving pulse. We attribute this effect to harmonic order-dependent dephasing, where doublon-holon dynamics for higher harmonics experience increased scattering from carriers injected throughout the laser pulse due to their longer trajectories [70,71]. Our study provides direct experimental observation of the strongly-driven sub-cycle doublon-holon dynamics, as well as the time- and energy-dependent dephasing within the femtosecond driving field. These results are important for understanding high-harmonic generation in quantum materials, correlated electronic effects in attosecond processes [69,70], and other related nonequilibrium phenomena such as light-induced phase transitions and Floquet engineering in quantum materials [74–77].

*hanzhe@purdue.edu

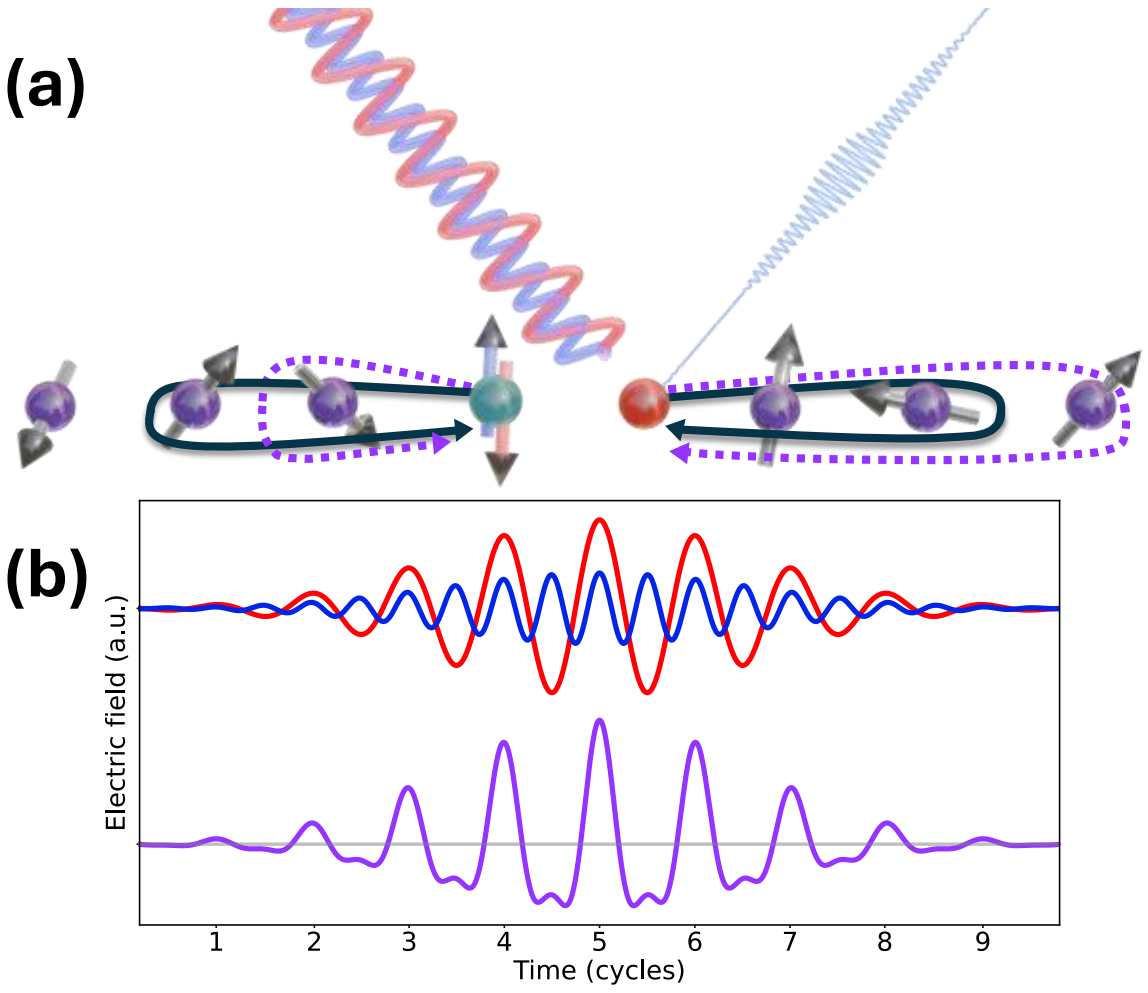

**Fig 1. Schematics of two-color driving. (a)** A phase-locked two-color field drives carrier tunneling across the Mott gap, creating doublon-holon (d-h) pairs. The field subsequently drives d-h motion through the correlated one-dimensional lattice. Recombination of the d-h pairs leads to the emission of high-harmonics. Black paths indicate symmetric excursions under fundamental-only driving, while purple dashed paths illustrate the asymmetric trajectories induced by the two-color field. **(b)** Electric field of the fundamental (red, top), second harmonic (blue, top), and total two-color field (purple, bottom). For a fundamental only excitation in an inversion-symmetric medium, opposite half cycles are equivalent, suppressing even-order harmonics. The interference of the fundamental with its second harmonic produces an asymmetry in the combined field whose sign alternates every half cycle. This asymmetric field breaks this half-cycle symmetry enabling even-harmonic emission.

We use a phase-locked two-color driving field to measure the high-harmonic chirp in $Ca_{0.4}Sr_{0.6}CuO_2$ (Fig. 1). Previously, this approach has been used to retrieve attochirp in atoms and semiconductors [33–35,42,48,49,78–80]. Its working principle is described in several prior works [33,34,49,79] and schematically illustrated in Fig. 1. Briefly, the driving field consists of an intense mid-infrared pulse superimposed with its weak second harmonic. The total oscillating laser field can become asymmetric, breaking inversion symmetry and generating even harmonics (Fig. 1b). Experimentally, even harmonic intensity is measured as a function of the two-color phase delay on sub-cycle timescales. High-harmonic chirp causes the even harmonics to be emitted at slightly different phases between the fundamental driving field and its second harmonic. Equivalently, by recording the relative phase delay in the two-color driving field for the peak position of each even harmonics, the attochirp and the sub-cycle electronic dynamics can be experimentally measured without using isolated attosecond pulses [81–83].

We choose single crystalline semi-one-dimensional $Ca_{0.4}Sr_{0.6}CuO_2$ Mott insulator as our test material. $Ca_{0.4}Sr_{0.6}CuO_2$ shares a similar crystal structure with the parent compound $SrCuO_2$, hosting pairs of Cu-O chains. With a weak inter-chain coupling [84–86], the electronic properties of $Ca_{0.4}Sr_{0.6}CuO_2$ are largely determined by the one-dimensional Cu-O chain. The spins in the Cu-O chain are disordered under room temperature. $Ca_{0.4}Sr_{0.6}CuO_2$ has a charge transfer gap of 2.1 eV. This material can be theoretically described using a one-dimensional Mott-Hubbard model with disordered spins [84–86].

For attochirp measurements, we set linearly polarized 3.5 μm and 1.75 μm two-color driving fields along the Cu-O chain. This driving photon energy (0.35 eV) is well below the charge transfer gap. The 3.5 μm excitation intensity is set at 0.6 TW/cm$^2$, below the damage threshold of the material. The weak 1.75 μm excitation is kept at a much lower intensity at 20 MW/cm$^2$. At such low intensity, the 1.75 μm beam only perturbs the 3.5 μm-driven high-harmonic generation process, without significantly modifying the underlying electronic dynamics nor generating harmonics alone. The sample is kept at room temperature. Under this condition, the 4$^{th}$ to 12$^{th}$ harmonics are collected in the reflection geometry. Figure 2 shows the experimental even high-harmonic spectral as a function of the two-color phase delay time. All above-gap even harmonics (6$^{th}$ to 12$^{th}$) show intensity modulation as a function of the two-color phase delay. The intensity modulates at every quarter of the 3.5 μm driving period, which is consistent with the symmetry-breaking two-color driving field in centrosymmetric materials [79]. Meanwhile, the odd harmonic intensity is largely unaffected by the two-color phase delay, confirming that the weak second harmonic does not directly modify the underlying high-harmonic generation process.

One key observation is that above-gap even harmonics (6$^{th}$ to 12$^{th}$) exhibit asynchronized intensity modulation as a function of two-color phase delay (Fig. 2, green label). This observation suggests that the above-gap harmonics are chirped, where different harmonics are emitted at different times within a laser driving cycle. Based on the peak modulation time of each harmonic (Fig. 2, green label), we extract a modulation chirp of 0.19 fs eV$^{-1}$ in this photon energy range. In contrast, the 4th harmonic, which is below the gap, does not show clear intensity modulation as a function of two-color phase.

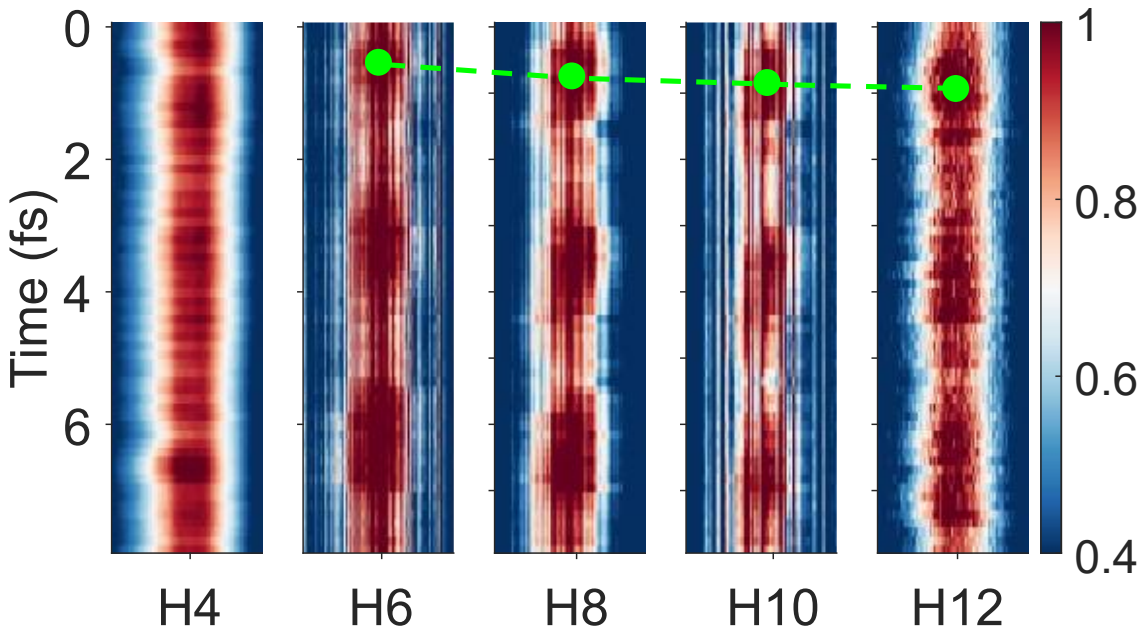

**Fig 2. Sub-cycle modulation of even-order harmonic emission under two-color driving.** Individually normalized spectra of the 4th, 6th, 8th, 10th, and 12th harmonics are shown as a function of the relative delay between the fundamental and second-harmonic fields. Each above-gap harmonic exhibits a modulation period of ~1/4 cycle of the fundamental. The green dashed line highlights the fitted delay of maximum harmonic yield, resulting in a measured harmonic chirp of 0.19 fs $eV^{-1}$.

The observed attochirp indicates that high-harmonic generation in $Ca_{0.4}Sr_{0.6}CuO_2$ originates from a recollision process. In the following section, we show that it is consistent with doublon-holon recombinations. Previous theoretical studies have attributed high-harmonic generation in Mott insulators to laser driven doublon-holon dynamics [59,60,62,66]. In the semiclassical model, each high harmonic is associated with a pair of doublon-holon trajectories. The harmonic photon energy is determined by the instantaneous doublon-holon energy at the time of recombination [58,59]. These recombination energies are associated with unique semiclassical doublon-holon trajectories and therefore map the recombination energy to a well-defined emission time. This procedure results in attochirp that depends on the doublon-holon dispersion. This picture is conceptually similar to the semiclassical electron-hole trajectories in conventional semiconductors [30,62].

This intuitive picture is supported by a semiclassical analysis of doublon-holon trajectories. Here, we perform a semiclassical analysis of doublon-holon trajectories, using laser and material parameters similar to the experimental conditions (Supplementary Materials). This approach was previously developed in Reference [62], which qualitatively agrees with the more sophisticated many-body calculations [62,67]. Briefly, the one-dimensional doublon-holon band dispersion is obtained using Bethe-ansatz (Fig. 3a). The band dispersion includes doublon-holon hopping and on-site Coulomb repulsion. Meanwhile, it ignores other many-body effects, such as spinon dynamics and non-local interactions, which spectrally broadens doublon-holon dispersion [62,87]. The semiclassical approach corresponds to the 'bare' doublon-holon dynamics, and can be considered as the lead-order contribution of a full many-body response [62,67,87].

The laser-driven semiclassical doublon-holon motion is iteratively solved in both reciprocal and real space. High harmonics only emit when doublons and holons meet in real space. The emitted photon energy equals the total doublon-holon energy at the recombination time (see [62] and Supplementary Materials).

Figure 3 b-d shows the semiclassical doublon-holon trajectory driven by a peak field of 0.21 V/ Å, which is similar to the experimental conditions. Figure 3b shows many recombining doublon-holon trajectories with positive emission chirps originating at different birth times in the laser driving cycle. With positive chirp, higher harmonics are emitted later in the optical cycle. Our simulation shows a positive chirp of 0.206 fs / eV, which is comparable to the experimental value of 0.19 fs / eV (Fig. 3d). This good agreement suggests that sub-cycle doublon-holon dynamics can be largely described by semiclassical trajectories in the correlated doublon-holon bands.

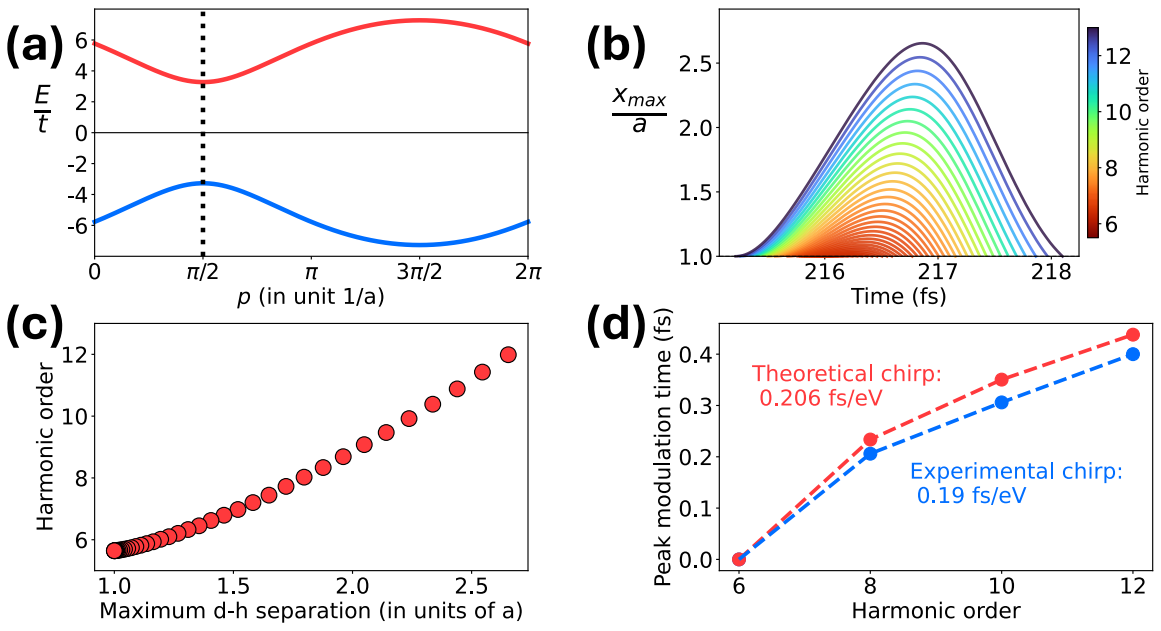

**Fig. 3. Semiclassical trajectory analysis of doublon-holon dynamics**. **(a)** Dispersions of the upper (red) and lower (blue) Hubbard bands obtained from Bethe ansatz. The vertical dashed line indicates the momentum of the minimum Mott gap. **(b)** Relative displacement of semiclassical doublon-holon trajectories as a function of time, color-coded by the harmonic order emitted upon recombination. **(c)** Emitted harmonic order as a function of the maximum doublon-holon separation distances. **(d)** Comparison of theoretically calculated (red) and experimentally measured (blue) chirp values. In this simulation, we choose electron hopping $t_h =$ 0.3 eV, on-site Coulomb repulsion U = 2.1 eV, and lattice constant $a$ = 3.9 Å. We use a 3.5 $\mu m$ driving pulse, with pulse duration 77.5 fs and peak field at 0.21 V/ Å.

These simulated trajectories are conceptually similar to the short trajectories in gas-phase high harmonics [33]. For noble gases, each harmonic is associated with two distinct electron trajectories. These trajectories have differing lengths and opposite chirps. They are termed as short and long trajectories. Like gas phase high-harmonic generation, long trajectories with negative chirp could in principle contribute to high harmonics in solids (Supplementary Material). However, they are not observed in our measurements. Our experiment only observes short trajectories with positive chirp. In this regime, higher-order harmonics originate from trajectories with larger doublon-holon spatial separations (Fig. 3c), which are more sensitive to dephasing due to increased carrier scattering events along the increased path length.

The lack of long trajectories could further suggest several dephasing mechanisms present on sub-cycle timescales. In band insulators, the suppression of long trajectories is largely attributed to the spatial spread of the electron wavepacket and scattering from other carriers and phonons [39,70,88]. This effect is important under strong field driving, where doublon-holon pairs could reach the edge of the Brillouin zone ($p = 3\pi/2$, Fig. 3a) and form Bloch oscillations within a single driving optical cycle. At the zone edge, the group velocity of doublons and holons changes direction, which can suppress recombination in long trajectories. Our simulation suggests that under experimental driving fields, doublon-holon pairs could form Bloch oscillations, leading to chaotic long trajectories with large recombination times (Supplementary Material). With extremely long doublon-holon excursion times, these trajectories are likely not to contribute to high-harmonic emission due to dephasing.

Beyond Bloch oscillations, many-body theories have predicted additional mechanisms, particularly doublon-holon hopping between weakly coupled Cu-O chains in similar compounds, could further suppress long trajectory contributions [67]. These processes are likely to coexist under experimental conditions. Separating these mechanisms requires further experimental control of inter-chain coupling through selective excitation or doping, and careful comparison between measured sub-cycle doublon-holon responses with numerical simulations in future studies.

We further show experimentally that doublon-holon dynamics experience ultrafast electronic dephasing that depends on the doublon-holon energies. To observe these effects, we measure the even high harmonic's temporal emission profile by scanning the relative time delay between 3.5 μm and 1.75 μm pulses. Figure 4a shows the measured temporal structure of even harmonics (4$^{th}$ to 12$^{th}$) as a function of the two-color phase delay, where early time indicates 1.75 μm arrives earlier relative to the 3.5 μm pulse. Apart from the sub-cycle oscillations, even harmonics only occur when the 3.5 μm and 1.75 μm pulse envelopes overlap (Fig. 4a). Importantly, as the harmonic order increases, the emission progressively shifts towards earlier cycles within the 3.5 μm driving pulse, leading to an asymmetric temporal emission profile. This effect is most clearly observed for the 12$^{th}$ harmonic (Fig. S3). The 12$^{th}$ harmonic reaches its intensity maximum ~ 22 fs earlier relative to the 4$^{th}$ harmonic and exhibits an asymmetric emission profile.

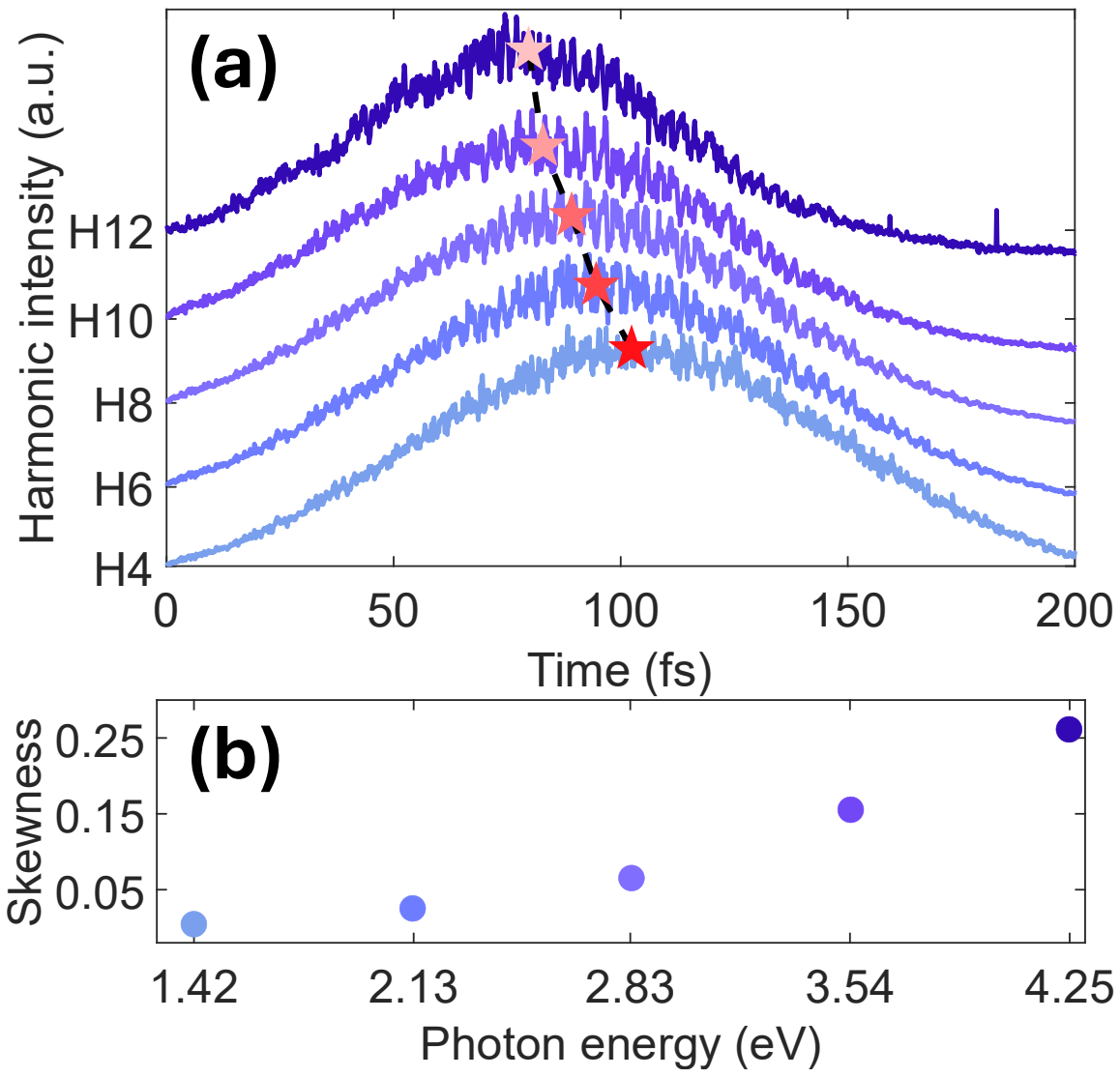

**Fig 4. Temporal evolution of harmonic emission. (a)** Integrated and individually normalized yield of even harmonics (4$^{th}$ to 12$^{th}$, labeled as H4-H12, respectively) as a function of two-color phase delay. The harmonic signals exhibit rapid quarter-cycle modulation within a slower emission envelope. Red stars mark the calculated envelope maxima. The envelope maximum shifts to earlier delay with increasing harmonic order. **(b)** Extracted temporal skewness of intensity distribution as a function of harmonic energy.

We use the intensity-weighted temporal skewness to quantify the asymmetry of the harmonic emission profile (Fig. 4b). The skewness increases monotonically with harmonic order, from 0.0045 for the 4$^{th}$ harmonic to 0.2613 for the 12$^{th}$ harmonic. The near-zero skewness of the below-gap 4$^{th}$ harmonic indicates a symmetric temporal response. The progressively positive skewness of the higher orders indicates that their emission becomes increasingly concentrated toward the earlier, lower-doublon-holon-population portion of the pulse.

We attribute the temporal emission profile shift to ultrafast electronic dephasing occurring within the multi-cycle, 3.5 μm driving pulse. As the laser driving cycle accumulates, the carrier population (doublons and holons) in the sample will gradually increase through tunneling or multiphoton excitations [66,89]. During the high-harmonic generation, the coherently driven doublon-holon pair will scatter from excess carriers in the background, causing dephasing between doublon-holon pairs and thus reducing the high-harmonic yield. Similar carrier-induced dephasing has been previously observed in band insulators, semiconductors, and semi-metals [70,71,83,90,91]. As electronic dephasing scales with the total carrier population, high harmonics are strongly suppressed in the later part of the driving pulse, where the total carrier population accumulates.

The harmonic order-dependent dephasing further supports the sub-cycle doublon-holon recombination dynamics. In this picture, the emitted high-harmonic photon energy equals the total doublon-holon pair energy at the instance of recombination. Typically, a longer doublon-holon trajectory results in higher order harmonics due to greater laser acceleration of the doublon-holon pair (Fig. 3c). Meanwhile, a longer doublon-holon trajectory is more sensitive to dephasing due to increased exposure to electronic scattering accumulated along its pathway, leading to greater dephasing for higher order harmonics.

Our trajectory analysis further provides an order-of-magnitude estimation of the characteristic dephasing length scale. Specifically, the semiclassical model suggests that doublon-holon travels ~ an additional 0.5 lattice site with increasing even harmonic orders (Fig. 3c) Meanwhile, the dephasing is quantitatively different between neighboring even harmonics, leading to modified temporal emission profile in the experiment (Fig. 4a). We thus conclude that under experimental conditions, electronic dephasing becomes important on a length scale of ~ 0.5 lattice sites, limited by scattering from back-ground doublon-holon pairs. It should be noted that this is an order-of-magnitude estimation of a characteristic length where doublon-holon dephasing becomes important to modify high-harmonic emission, limited by theoretical model accuracy. More accurate estimation of dephasing length can be in principle achieved by correlating similar measurements with full quantum many-body simulations.

Beyond high-harmonic generation, our experimental results give new insights into other nonequilibrium light-induced phenomena. One important example is Floquet engineering of emerging many-body states in solids. Floquet engineering requires long-wavelength periodic laser driving, with typical field strength approaching ~ 1 V/Å [75–77]. This condition is similar to typical solid-state high-harmonic generation. Our results suggest that under these conditions, coherent electronic responses could be suppressed within a single multi-cycle femtosecond driving pulse due to accumulated tunnel-ionized carriers from different optical cycles. Importantly, our experimental approach and similar multi-pulse wave-mixing experiments could be used for probing ultrafast coherence loss in periodically driven systems.

In conclusion, we report the experimental measurements of high harmonic chirp a semi-one-dimensional Mott insulator. The observed attochirp and harmonic order-dependent dephasing suggests that high-harmonic emission is dominated by doublon-holon recombination. The attochirp could be used as an experimental probe for sub-cycle doublon-holon motion under nonequilibrium conditions. Meanwhile, the harmonics' temporal emission profile could be used to quantify ultrafast, energy-dependent dephasing under intense, periodic driving conditions, which could be relevant for understanding the dephasing and thermalization bottlenecks for Floquet engineering of quantum matter [76].

We thank Philipp Werner and Yao Wang for insightful discussions. H.L. acknowledges support from the National Science Foundation under award number 2325212. L.H. acknowledges support from the NSF Graduate Research Fellowship Program. The work at Brookhaven National Laboratory was supported by the US Department of Energy, office of Basic Energy Sciences, contract no. DOE-SC0012704. Work at Los Alamos was supported by the Center for Integrated Nanotechnologies, a U.S. Department of Energy, Office of Basic Energy Sciences user facility. The authors disclose the use of Claude Opus 4.6 for assisting on python programming for simulation and data analysis. All AI-assisted scripts have been reviewed and validated by the authors.

# Supplementary Materials for

# Sub-cycle doublon-holon dynamics in one-dimensional Mott insulators revealed by two-color high-harmonic spectroscopy

Lance Hatch[1], Aditya Verma[1], Eric Schultz[2], Hanjun Yang[1], Priscila Rosa[3], Genda Gu[4], Igor Zaliznyak[4], Giulio Vampa[5,6], Laimei Nie[2], and Hanzhe Liu[1]

[1]Department of Chemistry, Purdue University, West Lafayette, Indiana 47907, USA

[2]Department of Physics and Astronomy, Purdue University, West Lafayette, Indiana 47907, USA

[3]Los Alamos National Laboratory, Los Alamos, New Mexico 87545, USA

[4]Department of Physics, Colorado State University, Fort Collins, CO 80523, USA

[5]Condensed Matter Physics and Materials Science Division, Brookhaven National Laboratory, Upton, New York 11973, USA

[6]Joint Attosecond Science Laboratory (JASLab), National Research Council of Canada and University of Ottawa, Ottawa, ON, Canada

## I. Experimental Details

Figures S1-S4.

## II. Theoretical Calculations

Figures S5-S6.

# I. EXPERIMENTAL DETAILS

## A. Experimental configuration

Experimentally, we use a 3.5 μm pulse as our fundamental driving field. Its second harmonic is generated through an AGS crystal. After AGS crystal, the co-propagating 3.5 μm and 1.75 μm beams are sent into a Mach-Zehnder interferometer. The interferometer allows precise control of the relative phase delay with attosecond accuracy. Additionally, the polarization and pulse energy of the two beams can be independently controlled.

## B. Raw harmonic spectra

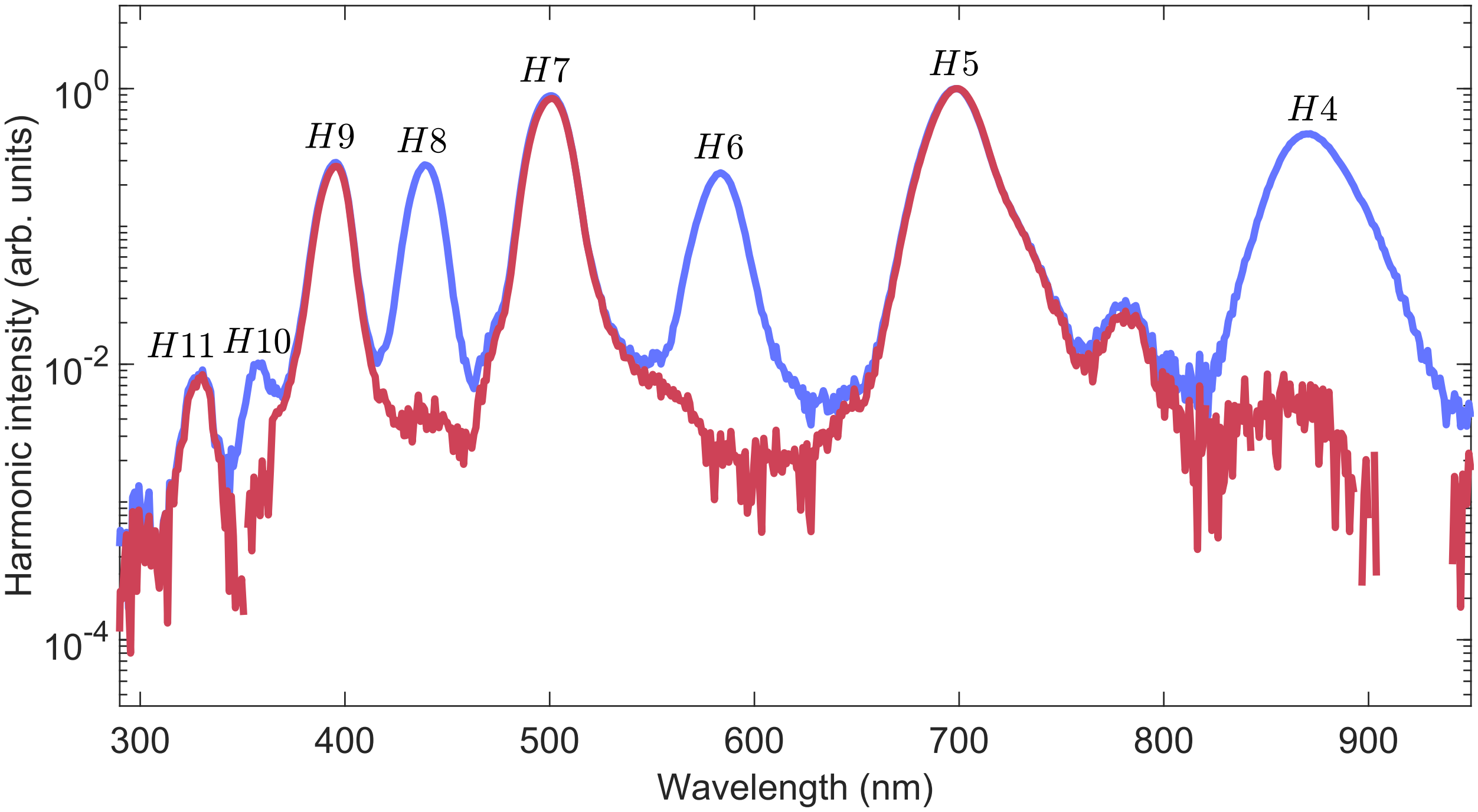

**Fig. S1: High-harmonic spectra of $Ca_{0.4}Sr_{0.6}CuO_2$ for 4th to 11th orders.** Harmonics are shown with single-color field (3.5 μm) driving (red) and a two-color field driving (blue). The addition of the second harmonic breaks symmetry, allowing for the generation of the even higher harmonics.

The 3.5 μm driving field was focused onto the sample using a 45 degree off-axis parabolic mirror. The high-harmonic radiation was collected in a reflection geometry and focused into a spectrometer (Princeton Instrument SP-2300i). The high-harmonic spectra were recorded by an electron multiplying charge-coupled device (EMCCD) detector (Princeton Instruments ProEM+ $1600^2$).

## C. Harmonic intensity dependence

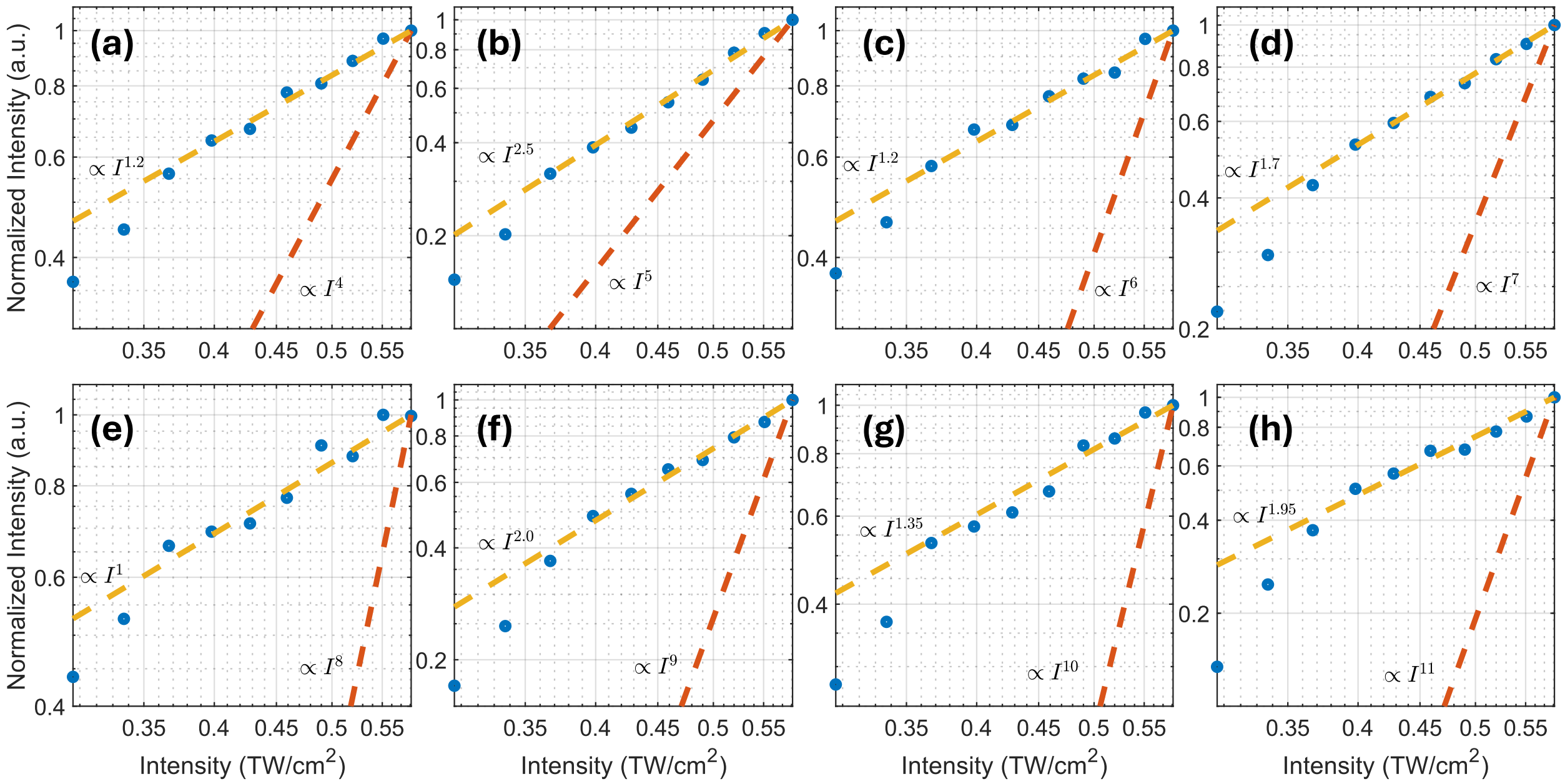

**Fig. S2: Dependence of HHG yields on fundamental field intensity. (a-h)** Log-normalized 4$^{th}$-11$^{th}$ harmonics, respectively. Blue circles denote the experimentally obtained data, yellow dashed lines are the fitted scaling parameters, and orange dashed lines represent the expected scaling from perturbation theory ($I_n \propto I^n_{MIR}$). Notably, all harmonic orders exhibit scaling that is typical of non-perturbative driving.

The pump power was controlled by two wire-grid polarizers (Thorlabs WP25M-IRA). The polarization of the pump pulse was controlled by a $MgF_2$ half-wave plate.

## D. Harmonic temporal profile

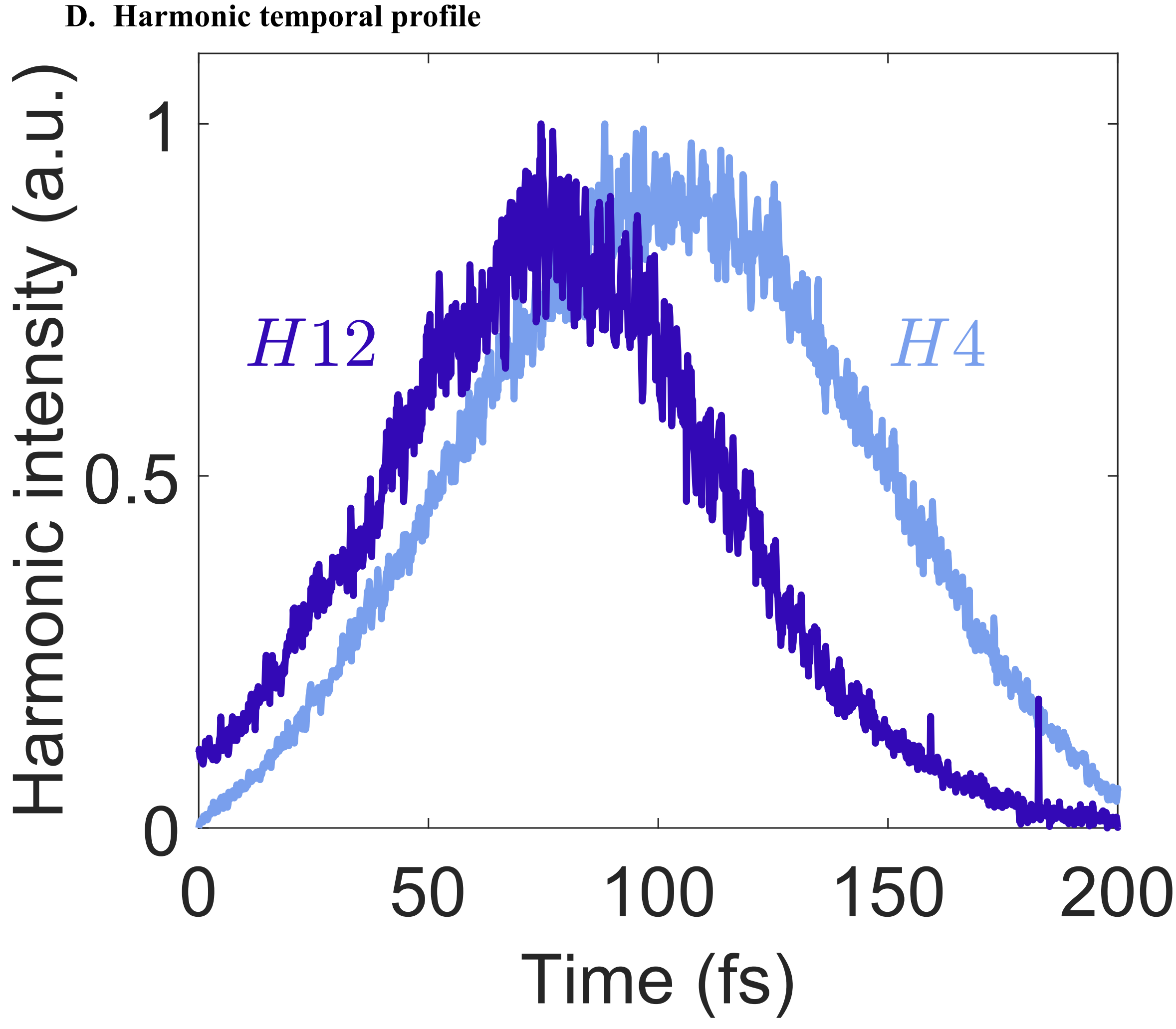

**Fig. S3: Comparison of temporal profile shift.** Independently normalized harmonic intensities as a function of two-color delay. The 12$^{th}$ harmonic peak (purple) is shifted noticeably to earlier in the cycle compared to 4$^{th}$ harmonic (light blue). Additionally, the reduction in the emission intensity is skewed for the 12$^{th}$-order after the peak in harmonic intensity relative to the 4$^{th}$-order.

## E. Calculation of temporal skewness

The temporal skewness of each harmonic emission profile was calculated by treating the measured harmonic intensity as a weight assigned to each experimental delay. For harmonic order $q$, the spectrally integrated intensity measured at delay $t_i$ is denoted by $I_q(t_i)$. The normalized temporal weight was defined as

$$p_{q,i} = \frac{I_q(t_i)}{\sum_i I_q(t_i)}$$

The intensity-weighted mean emission time was calculated as

$$\mu_q = \sum_i p_{q,i} t_i = \frac{\sum_i I_q(t_i) t_i}{\sum_i I_q(t_i)}$$

The corresponding intensity-weighted temporal variance was

$$\sigma_q^2 = \sum_i p_{q,i}(t_i - \mu_q)^2 = \frac{\sum_i I_q(t_i)(t_i - \mu_q)^2}{\sum_i I_q(t_i)}$$

The temporal skewness was then calculated from the standardized third central moment:

$$\gamma_q = \frac{\sum_i I_q(t_i)(t_i - \mu_q)^3}{[\sum_i I_q(t_i)]\sigma_q^3}$$

Equivalently, using normalized temporal weights,

$$\gamma_q = \frac{\sum_i p_{q,i}(t_i - \mu_q)^3}{\sigma_q^3}$$

Here, $\gamma_q$ is dimensionless and quantifies the asymmetry of the harmonic emission profile about its intensity-weighted mean. The value of $\gamma_q = 0$ corresponds to a symmetric temporal profile. Positive skewness indicates that the more extended temporal tail lies toward increasing experimental delay, whereas negative skewness indicates that the tail lies toward decreasing delay.

The calculated temporal skewness values were

$$\gamma_4 = 0.0045, \quad \gamma_6 = 0.0253, \quad \gamma_8 = 0.0653, \quad \gamma_{10} = 0.1554, \quad \gamma_{12} = 0.2613$$

The near-zero skewness of 4$^{th}$ harmonic indicates a symmetric temporal profile. In contrast, the increasingly positive skewness of 6$^{th}$ to 12$^{th}$ harmonic quantifies a monotonic increase in temporal asymmetry with harmonic order.

## F. Band gap determination

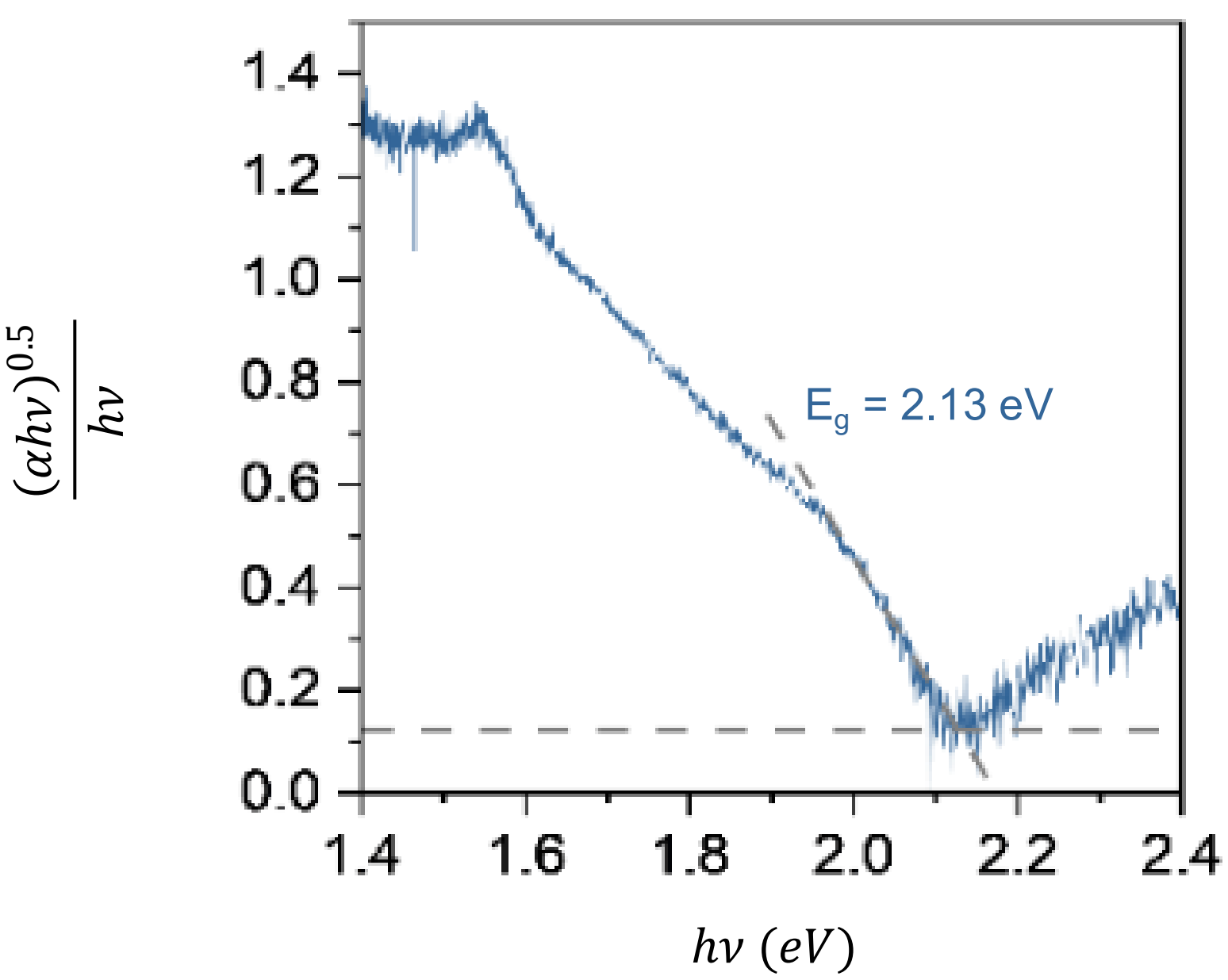

**Fig. S4: Optical band gap measurement.** Sample optical band gap is estimated by measuring its reflectivity using an Olympus BX53 microscope. A tauc plot based on indirect semiconductor was linearly extrapolated after Kubelka–Munk transformation, yielding an optical band gap of 2.13 eV.

# II. THEORETICAL CALCULATIONS

## A. Semiclassical doublon–holon trajectory model

To obtain a real-space picture of high-harmonic generation in the one-dimensional Mott insulator, we employed the semiclassical doublon-holon model introduced in Ref. [62]. This approach retains the Bethe-ansatz charge dispersion of the one-dimensional Hubbard model while treating the subsequent field-driven motion of the doublon-holon pair semi-classically. The model qualitatively reproduces the emission trajectories obtained from more complete many-body calculations [62], but neglects spinon dynamics, carrier scattering, and the back-action of the excited carriers on the driving field. It therefore provides a kinematic description of the trajectories contributing to each harmonic rather than a quantitative prediction of the harmonic intensity.

## Band structure and driving field

The doublon and holon dispersions were obtained from the Bethe-ansatz solution of the one-dimensional Hubbard model, following Ref. [62]. The total doublon-holon energy at relative crystal momentum $p$ is denoted by $E_{\mathrm{g}}(p)$. The upper and lower Hubbard bands shown in Fig. 4a are given by

$$E_{\pm}(p) = \pm \frac{E_{\mathrm{g}}(p)}{2}. \qquad \text{(S1)}$$

The calculation was performed using

$$\frac{U}{t_{\mathrm{h}}} = 10,$$

where $U$ is the on-site Coulomb interaction and $t_{\mathrm{h}}$is the nearest-neighbor hopping energy. Unless otherwise stated, the calculation uses normalized units in which

$$t_{\mathrm{h}} = \hbar = e = a = 1,$$

where $a$ is the lattice constant.

The driving laser was represented by the vector potential

$$A(t) = \frac{E_0}{\Omega} \exp\left[-\frac{(t - t_{\mathrm{c}})^2}{2\sigma^2}\right] \sin[\Omega(t - t_{\mathrm{c}})], \qquad \text{(S2)}$$

with electric field

$$\mathcal{E}(t) = -\frac{dA(t)}{dt}. \qquad \text{(S3)}$$

The parameters used in the calculation were

$$\Omega = 1.16,\ E_0 = 6.3,\ t_c = 100,\ \sigma = 30. \qquad \text{(S4)}$$

For hopping energy $t_h \approx 0.30$ eV, the driving photon energy is approximately 0.35 eV, corresponding to a wavelength of approximately 3.5 $\mu$m.

## Semiclassical propagation

The relative doublon-holon trajectory is governed by the semiclassical equations of motion [62],

$$\frac{dx(t)}{dt} = \left.\frac{dE_g(p)}{dp}\right|_{p=p(t)}, \qquad \text{(S5)}$$

and

$$p(t) = p(t_b) - A(t) + A(t_b), \qquad \text{(S6)}$$

where $x(t)$ is the spatial separation between the doublon and holon, $p(t)$ is their relative crystal momentum, and $t_b$is the pair-creation time.

The initial momentum was set to

$$p(t_b) = \frac{\pi}{2}, \qquad \text{(S7)}$$

corresponding to pair creation at the minimum Mott gap. The doublon and holon were initialized on neighboring lattice sites,

$$|\ x(t_b)\ | = a. \qquad \text{(S8)}$$

The sign of the initial spatial separation was chosen according to the direction of the electric field:

$$x(t_b) = \begin{cases} +a, & \mathcal{E}(t_b) > 0, \\ -a, & \mathcal{E}(t_b) < 0. \end{cases} \qquad \text{(S9)}$$

This convention ensures that the laser field initially drives the doublon and holon apart and produces reflection-symmetric trajectories for opposite field directions.

Trajectories were calculated for birth times between $t_b = 97.7$ and 99, with a spacing of approximately 0.002. The equations of motion were numerically integrated over $0 \le t \le 200$ using 100,000 time points. Of the 130 sampled birth times, 36 produced recombining trajectories within the simulated time window.

## Recombination and harmonic emission

A trajectory was considered to recombine when the pair returned inward to a nearest-neighbor separation after previously moving apart:

$$| x(t_{\mathrm{r}}) | = a, \qquad \text{(S10)}$$

where $t_{\mathrm{r}}$ is the recombination time. This criterion represents annihilation of the doublon–holon excitation through nearest-neighbor hopping.

The emitted photon energy was taken to be the total doublon–holon energy at recombination,

$$\hbar\omega_{\mathrm{emit}} = E_{\mathrm{g}}\,[p(t_{\mathrm{r}})]. \qquad \text{(S11)}$$

The corresponding harmonic order was calculated as

$$N_{\mathrm{H}} = \frac{\omega_{\mathrm{emit}}}{\Omega} = \frac{E_{\mathrm{g}}\,[p(t_{\mathrm{r}})]}{\hbar\Omega}. \qquad \text{(S12)}$$

The maximum spatial separation reached by each pair was defined as

$$x_{\max} = \max_{t_{\mathrm{b}} \le t \le t_{\mathrm{r}}} | x(t) |. \qquad \text{(S13)}$$

## Short and long trajectories

At low intensities (before the onset of Bloch oscillation), for each harmonic photon energy, there are two trajectories with different maximum separations and recombination times (Fig. S6 b, d). The two trajectories are termed as short and long trajectories. The short trajectory recombines earlier and exhibits a positive chirp,

$$\frac{d\omega_{\mathrm{emit}}}{dt_{\mathrm{r}}} > 0,$$

whereas the long trajectory recombines later and exhibits a negative chirp,

$$\frac{d\omega_{\mathrm{emit}}}{dt_{\mathrm{r}}} < 0.$$

These short and long trajectories are conceptually analogous to the short and long trajectories in gas-phase high harmonics. The experimentally observed positive chirp is therefore assigned to the short-trajectory branch. The absence of the predicted negative-chirp branch in our experiment is consistent with preferential suppression of the longer trajectories by dephasing. As the electrical field strength increases, doublon-holon pairs could be accelerated past the edge of the Brillouin zone. Upon passing the Brillouin zone (such as $p = 3\pi/2$), the doublon-holon velocity is reversed, resulting in Bloch oscillations. The associated trajectories exhibit chaotic oscillating behaviors, including extended trajectories and suppressed recombination (Fig S5b) in comparison to the low-field case (Fig S6b).

Along the short-trajectory branch, the maximum doublon-holon separation increases with harmonic order. Under experimental conditions, interpolation of the calculated trajectories gives

$$x_{\max}^{(6)} \approx 1.18a,\ x_{\max}^{(8)} \approx 1.80a,\ x_{\max}^{(10)} \approx 2.25a,\ x_{\max}^{(12)} \approx 2.65a. \qquad \text{(S14)}$$

Thus,

$$x_{\max}^{(8)} - x_{\max}^{(6)} \approx 0.62a, \qquad \text{(S15)}$$
$$x_{\max}^{(10)} - x_{\max}^{(8)} \approx 0.45a, \qquad \text{(S16)}$$
$$x_{\max}^{(12)} - x_{\max}^{(10)} \approx 0.40a, \qquad \text{(S17)}$$

where $x_{\max}^{(i)}$ is the max doublon-holon spatial separation for the $i$-th harmonic. Increasing the even harmonic order by two therefore corresponds to an additional propagation distance of approximately 0.5 lattice sites.

The semiclassical model does not explicitly calculate dephasing. Nevertheless, when combined with the experimentally observed difference in dephasing between neighboring even harmonics, the calculated change in propagation distance suggests that electronic dephasing becomes important over a characteristic distance of approximately 0.5 lattice sites. This value should be regarded as an order-of-magnitude estimate rather than a directly calculated microscopic mean free path.

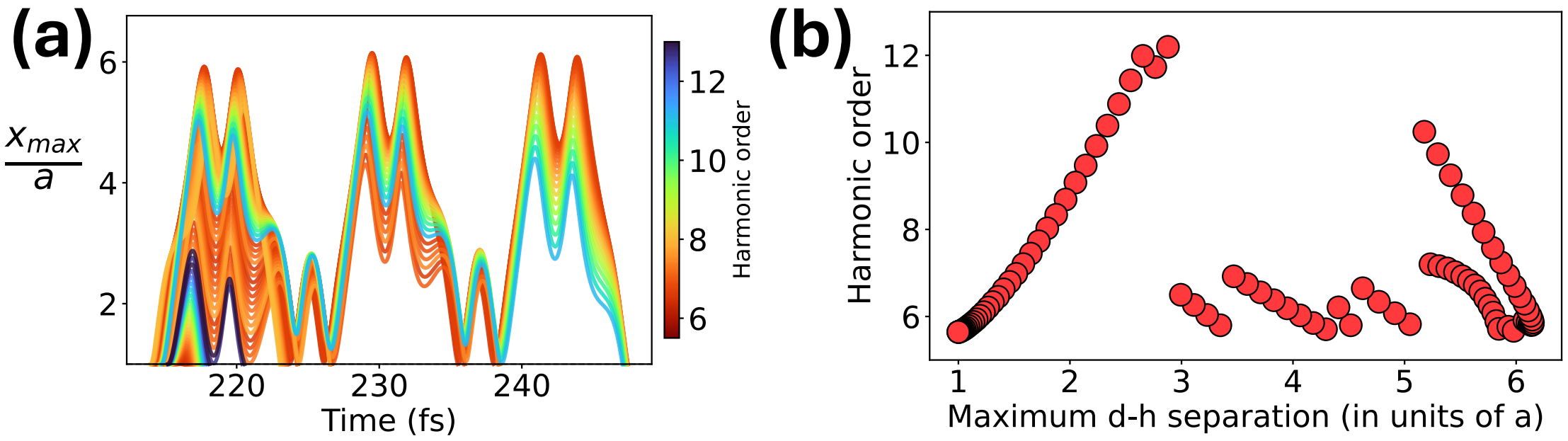

**Fig. S5: Semiclassical trajectory analysis of doublon-holon dynamics under a high driving field (0.21 V/Å). (a)** Relative displacement of semiclassical doublon-holon trajectories as a function of time, color-coded by the harmonic order emitted upon recombination. We only show recombining trajectories within a birth time window between 213.96 and 216.81 fs. In this plot, we set the doublon-holon propagation time up to 37 fs. Under this condition, Bloch oscillation leads to chaotic and extended long trajectories, which are not expected to contribute to high-harmonic emission due to dephasing. **(b)** Emitted harmonic order as a function of the maximum doublon-holon separation distances. The doublon-holon separation for well-defined short trajectories is limited to below ~ 3 lattice sites.

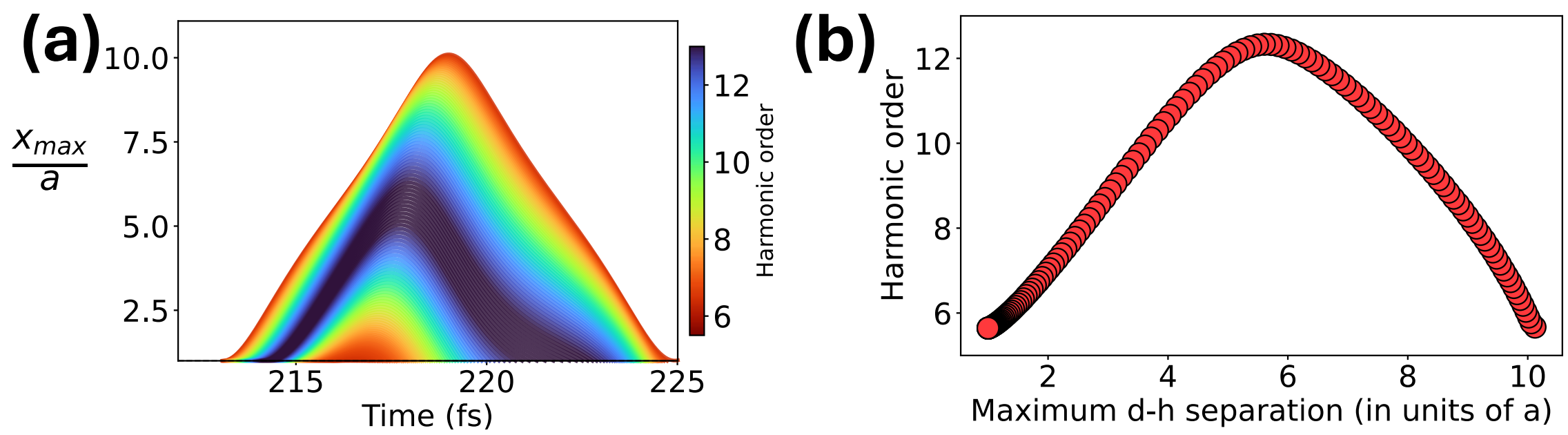

**Fig. S6: Semiclassical trajectory analysis of doublon-holon dynamics under a low field driving (0.0525 V/Å). (a)** Relative displacement of semiclassical doublon-holon trajectories as a function of time, color-coded by the harmonic order emitted upon recombination. Under this weak driving field condition, doublon-holon pairs do not form chaotic Bloch oscillations, leading to well-defined long- and short trajectories. **(b)** Emitted harmonic order as a function of the maximum doublon-holon separation distances. With weaker field, the maximum doublon-holon spatial separation is larger.

## Calculation of even harmonic modulation as a function of two-color phase

Based on simulated doublon-holon trajectories, we calculate even harmonic intensity modulation as a function of two-color phase and compare with our experiment. Similar analysis has been previously applied to noble gases and band insulators, which are described in Ref. 33 and 34. In this description, doublons and holons acquire an additional phase from the weak second harmonic perturbation. For each doublon-holon trajectory, this additional accumulated phase is written as:

$$\sigma(\varphi) = \int_{t_{birth}}^{t_{recombine}} dt\, \boldsymbol{v}(t) \cdot \boldsymbol{A}_{2w}(\varphi)\ ,$$

where $\boldsymbol{v}(t)$ is the doublon (or holon) speed along the trajectory. $\boldsymbol{A}_{2w}(t,\varphi) = A_2 sin(2\omega t + \varphi)$ is the vector potential of the weak second harmonic at 1.75 $\mu$m. $\varphi$ is the relative phase between the fundamental 3.5 $\mu$m and second harmonic 1.75 $\mu$m field. We set the second harmonic peak field at ~173 times smaller compared to the fundamental field, which is comparable to the experimental conditions.

Generally, $\sigma(\varphi)$ is a periodic function in phase $\varphi$, and can be written as $\sigma(\varphi) \sim \cos(\varphi - \theta)$. The experimentally measured even harmonic intensity is proportional to $|\cos(\varphi - \theta)|^2$ (Ref. 33), which is maximized at two-color driving field phase $\varphi_{max} = \theta$. For each trajectory plotted in Figure 3b, we calculate accumulated phase $\sigma(\varphi)$ to extract parameter $\theta$ for each trajectory. As each trajectory is associated with a unique emitted harmonic photon energy, the two-color phase $\varphi_{max} = \theta$ that maximizes each even harmonics can be extracted and compared with the experimental results, as shown in Figure 3d.